\documentclass[conference]{IEEEtran}
\IEEEoverridecommandlockouts
\usepackage{cite}
\usepackage[hidelinks]{hyperref}
\usepackage{amsmath,amssymb,amsfonts}
\usepackage{algorithmic}
\usepackage{graphicx}
\usepackage{textcomp}
\usepackage{array}
\usepackage{lipsum} 
\usepackage{listings}
\usepackage[table,xcdraw]{xcolor}
\usepackage{url}
\def\BibTeX{{\rm B\kern-.05em{\sc i\kern-.025em b}\kern-.08em
    T\kern-.1667em\lower.7ex\hbox{E}\kern-.125emX}}

\usepackage{orcidlink}

\begin{document}

\title{Optimal Planning for Enhancing the Resilience of Modern Distribution Systems Against Cyberattacks\\
}

\author{\IEEEauthorblockN{Armita Khashayardoost\orcidlink{0009-0005-9123-368X}}
\IEEEauthorblockA{\textit{Department of Engineering Science} \\
\textit{University of Toronto}\\
Toronto, ON, Canada \\
a.khashayardoost@mail.utoronto.ca}
\and
\IEEEauthorblockN{Ahmad Mohammad Saber\orcidlink{0000-0003-3115-2384}}
\IEEEauthorblockA{\textit{ECE Department} \\
\textit{University of Toronto}\\
Toronto, ON, Canada \\
ahmad.m.saber@ieee.org}
\and
\IEEEauthorblockN{Deepa Kundur\orcidlink{0000-0001-5999-1847}}
\IEEEauthorblockA{\textit{ECE Department} \\
\textit{University of Toronto}\\
Toronto, ON, Canada \\
dkundur@ece.utoronto.ca}
}

\maketitle




\begin{abstract}
The increasing integration of IoT-connected devices in smart grids has introduced new vulnerabilities at the distribution level. Of particular concern is the potential for cyberattacks that exploit high-wattage IoT devices, such as EV chargers, to manipulate local demand and destabilize the grid. While previous studies have primarily focused on such attacks at the transmission level, this paper investigates their feasibility and impact at the distribution level. We examine how cyberattackers can target voltage-sensitive nodes, especially those exposed by the presence of high-consumption devices, to cause voltage deviation and service disruption. Our analysis demonstrates that conventional grid protections are insufficient against these intelligent, localized attacks. To address this, we propose resilience strategies using distributed generation (DGs), exploring their role in preemptive planning. This research highlights the urgent need for distribution-level cyber resilience planning in 
smart grids.
\end{abstract}

\begin{IEEEkeywords}
Cybersecurity, Distributed Generation, IoT, Load-Altering Attacks, Power, Distribution Systems, Resilience.
\end{IEEEkeywords}

\section{Introduction}
Smart grids represent the next generation of electricity management, integrating advanced technologies with energy distribution to enhance efficiency, reliability, and sustainability \cite{madiot2.0}. Recent studies have shown that cyberattacks can target critical nodes in smart grid transmission systems and bypass traditional protection measures, potentially leading to large-scale blackouts and service disruptions \cite{207377}. With the increasing prevalence of electric vehicle (EV) chargers, smart home technologies, and other high-wattage technologies, the distribution level of the grid is becoming particularly vulnerable \cite{khan2019impact}.
These devices, often referred to as Internet of Things (IoT) devices, are capable of communicating and exchanging data over networks. When compromised, they can become load-altering tools for cyberattackers, creating critical nodes, points in the distribution system with low voltage magnitudes that are particularly susceptible to voltage deviations \cite{madiot1.0, 6299002}. Such deviations can destabilize the entire voltage profile of a distribution network.

Most existing research has focused on the impact and mitigation of these types of cyberattacks at the transmission level \cite{madiot1.0, madiot2.0}. However, the distribution network has received significantly less attention, despite being more exposed. Distribution systems operate at lower voltage levels, typically follow radial topologies, and involve a higher degree of interaction with end-users \cite{madiot1.0}. They also rely on less sophisticated control mechanisms and have limited built-in security and monitoring infrastructure \cite{power2017}, making them more susceptible to cyberattacks, especially as IoT device penetration increases.

This research aims to address that gap by analyzing how high-wattage IoT devices, when compromised, can be used to launch targeted attacks on vulnerable points within a distribution network. We further explore the potential for distributed generation (DG) to serve as preventive resilience strategies. Our key goals are to:
\begin{enumerate}
    \item demonstrate the severity of IoT-based cyberattacks at the distribution level; and
    \item evaluate whether strategic placement of DGs can reduce vulnerability and improve system resilience post-attack.
\end{enumerate}

To accomplish this, we employ the IEEE 33-Bus Distribution Network as a testbed and use the Manipulation of Demand via IoT (MaDIoT) 2.0 attack framework as the basis for our cyber threat model \cite{ madiot2.0}. After establishing system vulnerabilities, we develop criteria for strategic DG placement: focusing on minimizing system power loss, improving voltage profiles, reducing investment costs, and maximizing DG output \cite{6299002}. We then re-simulate the attack scenario to assess performance under these improved conditions. This work contributes to the growing field of grid cybersecurity by highlighting vulnerabilities at the distribution level and proposing tangible mitigation strategies to ensure continuity and stability in increasingly digitized energy systems.

\section{Vulnerabilities of Modern Distribution Systems to Cyberattacks}

\begin{figure}[t!]
    \centering
    \includegraphics[width=0.48\textwidth]{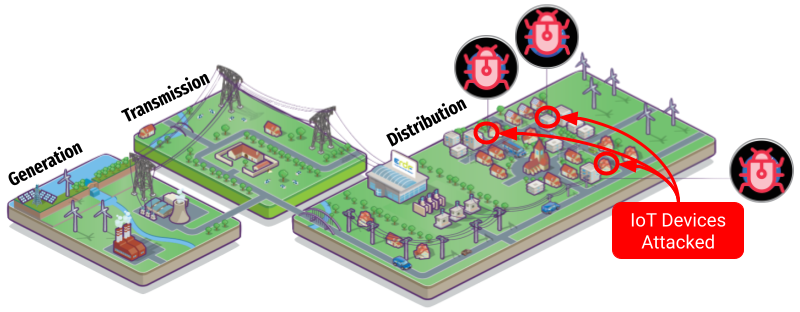}
    \caption{Illustration of IoT-based attacks on distribution systems}
    \label{fig1}
\end{figure}

Distribution systems are particularly vulnerable to cyberattacks in modern power networks, especially compared to the generation and transmission layers \cite{countermadiot}. 
Historically, distribution networks were not also designed with cybersecurity in mind.
Operating at lower voltages, they are also more sensitive to demand fluctuations and component failures. Furthermore, the growing integration of IoT-enabled devices at the grid's edge has significantly expanded the potential attack surface. As illustrated in Figure \ref{fig1}, adversaries can exploit these vulnerabilities by targeting edge devices, potentially disrupting localized service and destabilizing broader system operations \cite{power2017, ieso2025}.

Maintaining a stable grid requires a real-time balance between power supply and demand, as electricity cannot be stored at scale \cite{energygov2023}. Sudden shifts in load, whether from increased consumption or generation outages, can cause frequency and voltage deviations, triggering automatic protections, equipment failures, or even large-scale blackouts \cite{power2017}. As this research explores, these dynamics can be weaponized through cyber threats.

\subsection{Cyber Vulnerabilities in Distribution-level IoT Devices}

Smart grids increasingly rely on IoT devices, such as EV chargers, smart thermostats, and heat pumps, to manage energy use and automate grid responses. However, these devices often lack strong security protocols, with many transmitting unencrypted data and using weak or default credentials \cite{iotresearch}. A 2022 report found that 98\% of IoT traffic is unencrypted, leaving devices open to exploitation \cite{consumersmartdevices}. Attackers can gain access to IoT networks and send load-altering commands, manipulating demand in targeted ways that disrupt grid stability. Rather than disabling the infrastructure, they can subtly destabilize it, triggering voltage drops, frequency deviations, or even cascading failures. What makes this even more dangerous is the openness of modern electricity markets. To promote transparency and competition, operators publicly release real-time data on electricity pricing, demand, and congestion. This gives attackers the tools to map out weak points in the grid, identifying areas under stress or heavy load, and time their attacks for maximum disruption \cite{marketresearch}. Together, the widespread insecurity of IoT devices and the public availability of granular grid data create a perfect storm: attackers can not only infiltrate, but also strategically plan where and when to strike for the greatest impact.

\subsection{MaDIoT Attacks and Grid-Level Threats}

MaDIoT attacks exploit compromised smart devices to coordinate demand spikes that destabilize the grid. Originally introduced by Soltan et al. \cite{madiot1.0}, MaDIoT 1.0 attacks simulate sudden load changes, leading to frequency imbalances, line overloads, and even blackouts. Table \ref{tab:appliances} shows the power usage of some vulnerable devices, highlighting their disruptive potential. Soltan’s simulations included five attack scenarios, from frequency manipulation to increasing peak-hour operating costs, and showed how even a 30\% increase in demand could trip all generators in the Western U.S. grid. A 1\% increase caused 263 line failures in the Polish grid, cutting 86\% of its load. However, the study dismissed the role of distribution systems, claiming they were not significantly affected. Huang et al. \cite{countermadiot} challenged this, noting that Soltan’s model ignored real-world protection mechanisms, like generator protection, UFLS, and voltage or overcurrent relays. When included in simulations, these protections prevented blackouts, even with 8 million compromised devices, by partitioning the system into smaller, self-sustaining “islands.” While this reduced the likelihood of full blackouts, it still left isolated regions vulnerable to future contingencies. Crucially, Huang acknowledged that distribution systems lack many of these protections, making them easier to compromise. Yet, like Soltan, their study did not explore distribution-level risks in detail. Shekari et al. \cite{madiot2.0} extended the work with MaDIoT 2.0, a more targeted attack that uses grid topology, voltage stability indices, and modal analysis to pinpoint weak spots. Simulations on the IEEE-39 and IEEE-9 test systems showed success rates of 91\% and 67\%, respectively, using as few as 150,000 bots, a dramatic drop from the millions previously needed. These refined attacks are harder to detect, as they mimic natural load variation and bypass traditional protection mechanisms.

\begin{table}[t!]
    \centering
    \caption{Home appliances’ approximate electric power usage based on appliances manufactured by General Electric \cite{geappliances2025} and data from the Department of Energy \cite{energygov2017}}
    \label{tab:appliances}
    \fontsize{10}{12}\selectfont
    \begin{tabular}{|c|c|}
        \hline
        \rowcolor{gray!30} \textbf{Appliances} & \textbf{Power Usage [W]} \\
        \hline
        Air Conditioner & 1000 \\
        \hline
        Space Heater & 1500 \\
        \hline
        Electric Water Heater & 5000 \\
        \hline
        EV Charger & 7000 \\
        \hline
    \end{tabular}
\end{table}

Despite highlighting the threat, all three papers focused solely on transmission systems and offered limited actionable mitigation. This is a major oversight: the distribution grid, where most IoT devices reside, is less protected and equally critical. If exploited, distribution-level MaDIoT 2.0 attacks could achieve similar disruptions, bypassing transmission-focused defenses entirely.

\section{Security Evaluation Framework}

This section outlines the analytical foundations used to evaluate node vulnerability and optimize DG placement for resilience enhancement against cyber-physical attacks in smart distribution grids.

\subsection{Voltage Sensitivity-Based Node Criticality}
Node criticality is assessed through Voltage Magnitude (VM) analysis, which identifies the buses with the lowest voltage magnitudes under normal conditions. These nodes are considered most susceptible to voltage 
violation or unnecessary disconnections (hence supply interruptions)
when subjected to additional load and are therefore prioritized as critical. The VM index for a bus \( i \) is defined simply as:
\begin{equation}
    VM_i = V_i
\end{equation}
where \( V_i \) is the steady-state voltage magnitude at bus \( i \). Lower \( VM_i \) values indicate higher vulnerability, as they reflect weaker voltage support and greater sensitivity to perturbations. 

The VM method offers a computationally efficient means of critical node identification by ranking buses based on voltage magnitude, avoiding the need for complex modal or eigenvalue-based analysis. While alternative approaches such as the Voltage Stability Index (VSI), Fast Voltage Stability Index (FVSI), and modal analysis can provide higher fidelity, they are significantly more computationally intensive and less scalable for real-time or large-system applications. As demonstrated by Shekari et al. \cite{madiot2.0}, the VM method achieves a favorable balance between simplicity and effectiveness in cyberattack simulations, and it accurately highlights the correlation between low voltage magnitude and node criticality.

\subsection{Resilience Through Distributed Generation}
\label{section1}

The integration of DG units is proposed as a resilience strategy to counter voltage instability induced by cyberattacks. Strategically placed DGs can locally support voltage, reduce losses, and improve system stability.

The DG placement and sizing problem is formulated as a multi-objective optimization problem, minimizing voltage deviation and system losses:
\begin{equation}
    \min \quad \sum_{i=1}^{N} \left( |V_i - V_{\text{ref}}|^2 + \alpha \cdot P_{\text{loss}}(i) + \beta \cdot Q_{\text{loss}}(i) \right)
\end{equation} \cite{ieee_eq}
where \(V_i\) is the voltage at each bus, \(V_{\text{ref}}\) is the reference voltage, \(P_{\text{loss}}(i)\) represents real power losses, and \(Q_{\text{loss}}(i)\) represents reactive power losses. The weighting coefficients \(\alpha\) and \(\beta\) are used to balance voltage regulation with power loss minimization.
The constraints on the DG placement are given by:
\begin{equation}
    P_{\text{DG}, \min} \leq P_{\text{DG},i} \leq P_{\text{DG}, \max}, \quad \text{for each } i = 1, 2, 3
\end{equation}
where \(P_{\text{DG},i}\) represents the power generated by DG at node \(i\), and \(P_{\text{DG}, \min} = 100 \, \text{kW}\), \(P_{\text{DG}, \max} = 5000 \, \text{kW}\).
DGs 
can be assumed to operate at unity power factor. The goal is to enhance both operational efficiency and cyber-resilience.

Particle Swarm Optimization (PSO) is employed
in this paper as a tool to solve the DG placement problem. PSO simulates social behavior in populations to explore optimal solutions without requiring gradient information. Each particle updates its velocity and position based on its own and its neighbors' best-known positions. The benefits of PSO include: 1) strong global search capabilities in non-convex spaces; 2) flexibility for nonlinear, multi-objective problems, and 3) efficient convergence with modest computational resources.

\section{Simulation Setup}

\begin{figure}[t!]
    \centering
    \includegraphics[width=0.5\textwidth]{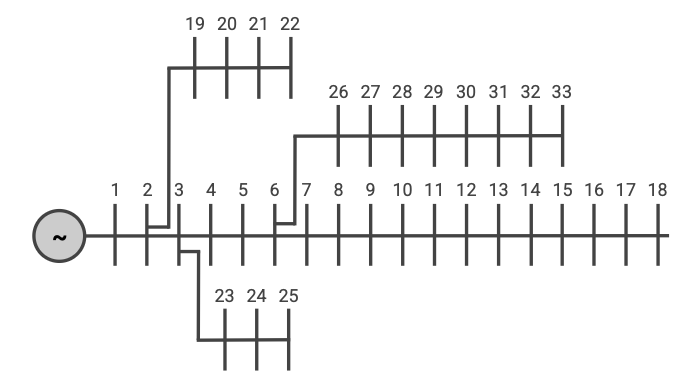}
    \caption{The IEEE 33-bus distribution system}
    \label{ieee33}
\end{figure}

Simulations are conducted on the IEEE 33-bus radial distribution system, shown in Figure~\ref{ieee33}. This system was chosen for its simplicity, branch-based layout, and widespread use in distribution system research, which ensures compatibility with existing literature and tools.

Voltage magnitudes are computed under normal operating conditions to establish a steady-state baseline. The allowable voltage range is set between 0.917 and 1.042 per unit (p.u.), as recommended for single-phase distributed systems~\cite{tonkoski2012impact}, and critical buses are identified via VM analysis.

To simulate cyberattacks, controlled load fluctuations ranging from ±5\% to ±15\% are introduced at three critical buses. These buses are selected based on their low voltage magnitudes, which indicate high sensitivity to load disturbances. Attack scenarios include both step-based and oscillating load alterations. The system’s response to these adversarial load variations is monitored, with a particular focus on whether bus voltages remain within the allowable range. Violations of this range are used to assess the system's inherent vulnerability to cyber-induced instabilities.

DG units are integrated into the system using the PSO-based optimization framework described in Section~\ref{section1}. A swarm size of 500 particles is used to ensure solution diversity while maintaining convergence stability. A voltage weight of 0.5 and a total loss weight of 0.5 are applied, with the loss component equally divided between real and reactive losses. This yields \(\alpha\) = 0.25 for real power losses and \(\beta\) = 0.25 for reactive power losses. Optimal locations and sizes are determined based on voltage support and power loss minimization. After integration, the same set of load-altering attacks is re-applied to evaluate improvements in system resilience. Voltage magnitudes are monitored again across all buses and compared against the base case results.

This post-mitigation analysis serves two purposes: (1) to quantify the extent of voltage stabilization provided by the DGs, and (2) to assess whether new vulnerabilities are introduced through DG integration. The comparison allows for a more nuanced understanding of the trade-offs and effectiveness of DG-based mitigation strategies in the context of cyber-physical grid security.

Overall, three case studies are analyzed:
\begin{enumerate}
    \item Baseline: Step-based load-altering attacks with no DGs.
    \item DG Step Attack: Step-based load-altering attacks with DGs integrated.
    \item DG Dynamic Attack: Oscillating, time-varying attacks with DGs integrated.
\end{enumerate}

This framework enables the assessment of both the impact of cyberattacks and the effectiveness of the proposed resilience strategy. This study investigates the vulnerability of distribution systems to coordinated load-altering attacks and examines the extent to which DGs can mitigate their impact.

\section{Results}
This section applies the methods to assess the resilience of smart grid under cyberattacks, focusing on how DGs influences grid stability, including both step-attack and dynamic stability.

\subsection{Case Study 1: Distribution System with No DGs}
\begin{figure}[t!]
\centerline{\includegraphics[scale=0.25]{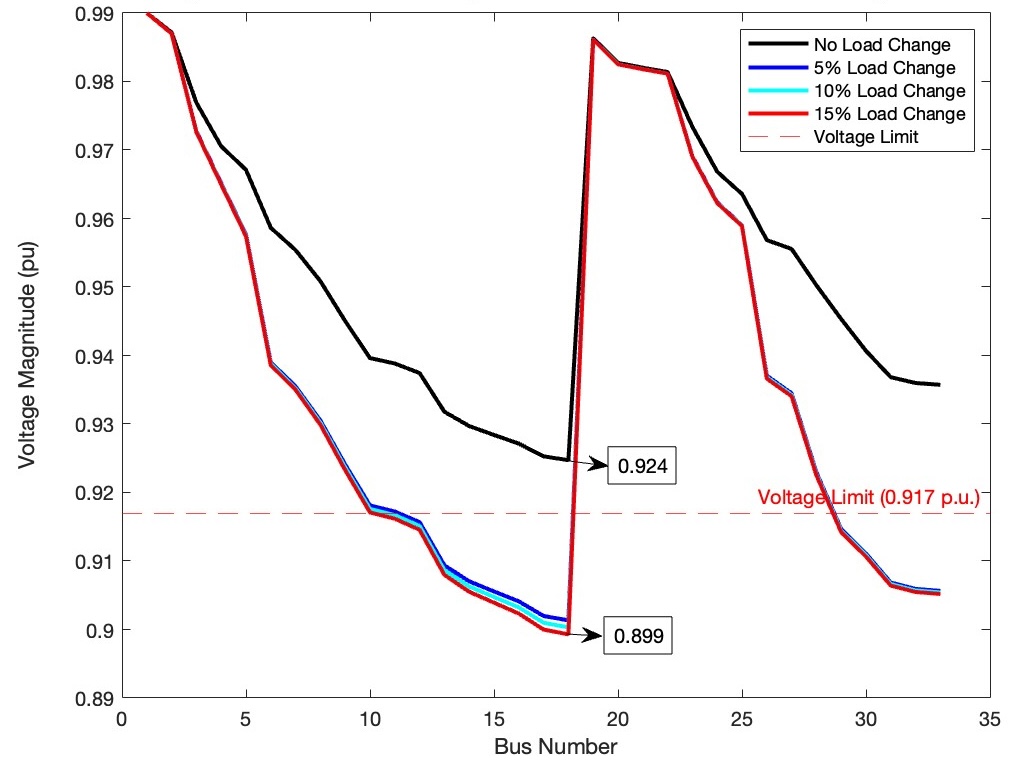}}
\caption{Voltage profile for base case with load-altering attacks}
\label{base_attack}
\end{figure}

In the IEEE 33-bus system without DGs, the voltage magnitudes under normal conditions are found to range from 0.926 to 0.987 p.u. shown in Black in Figure \ref{base_attack}, well within the standard operating range \cite{tonkoski2012impact}. We then use VM analysis to identify critical nodes that are prone to voltage instability under cyberattack conditions. As shown in Table \ref{tab:base_VM}, the buses with the lowest voltage magnitudes are Buses 16, 17, and 18. These buses are selected as targets for the load-altering cyberattacks, with results summarized in Table \ref{tab:my-attack_power}. It is clear that the impact of a 15\% load increase on both real and reactive power losses is significant, with the real power loss increasing by 52\% and the reactive power loss by 51\%. The results indicate that even small increases in load at these critical nodes can result in large increases in both real and reactive power losses, signaling a deterioration of the system's stability under attack. This demonstrates the susceptibility of the grid to cyber-induced power losses, particularly in scenarios where load disruptions are targeted at already weak points.
\begin{table}[t!]
    \centering
    \caption{VM for IEEE 33-bus base case system where red represents the lowest voltage values}
    \label{tab:base_VM}
    \resizebox{\columnwidth}{!}{%
    \begin{tabular}{|c|c|cc}
    \hline
    \rowcolor{gray!30} 
    \textbf{Bus} & \textbf{Voltage Magnitude} & \multicolumn{1}{c|}{\textbf{Bus}} & \multicolumn{1}{c|}{\textbf{Voltage Magnitude}} \\ \hline
    \textbf{1} & 0.9900 & \multicolumn{1}{c|}{\textbf{18}} & \multicolumn{1}{c|}{\cellcolor[HTML]{F4CCCC}0.9247} \\ \hline
    \textbf{2} & 0.9872 & \multicolumn{1}{c|}{\textbf{19}} & \multicolumn{1}{c|}{0.9863} \\ \hline
    \textbf{3} & 0.9769 & \multicolumn{1}{c|}{\textbf{20}} & \multicolumn{1}{c|}{0.9827} \\ \hline
    \textbf{4} & 0.9705 & \multicolumn{1}{c|}{\textbf{21}} & \multicolumn{1}{c|}{0.9820} \\ \hline
    \textbf{5} & 0.9671 & \multicolumn{1}{c|}{\textbf{22}} & \multicolumn{1}{c|}{0.9814} \\ \hline
    \textbf{6} & 0.9586 & \multicolumn{1}{c|}{\textbf{23}} & \multicolumn{1}{c|}{0.9734} \\ \hline
    \textbf{7} & 0.9554 & \multicolumn{1}{c|}{\textbf{24}} & \multicolumn{1}{c|}{0.9668} \\ \hline
    \textbf{8} & 0.9509 & \multicolumn{1}{c|}{\textbf{25}} & \multicolumn{1}{c|}{0.9636} \\ \hline
    \textbf{9} & 0.9450 & \multicolumn{1}{c|}{\textbf{26}} & \multicolumn{1}{c|}{0.9568} \\ \hline
    \textbf{10} & 0.9396 & \multicolumn{1}{c|}{\textbf{27}} & \multicolumn{1}{c|}{0.9556} \\ \hline
    \textbf{11} & 0.9388 & \multicolumn{1}{c|}{\textbf{28}} & \multicolumn{1}{c|}{0.9503} \\ \hline
    \textbf{12} & 0.9374 & \multicolumn{1}{c|}{\textbf{29}} & \multicolumn{1}{c|}{0.9453} \\ \hline
    \textbf{13} & 0.9318 & \multicolumn{1}{c|}{\textbf{30}} & \multicolumn{1}{c|}{0.9407} \\ \hline
    \textbf{14} & 0.9297 & \multicolumn{1}{c|}{\textbf{31}} & \multicolumn{1}{c|}{0.9368} \\ \hline
    \textbf{15} & 0.9284 & \multicolumn{1}{c|}{\textbf{32}} & \multicolumn{1}{c|}{0.9360} \\ \hline
    \textbf{16} & \cellcolor[HTML]{F4CCCC}0.9271 & \multicolumn{1}{c|}{\textbf{33}} & \multicolumn{1}{c|}{0.9357} \\ \hline
    \textbf{17} & \cellcolor[HTML]{F4CCCC}0.9253 & \multicolumn{2}{c}{} \\ \cline{1-2}
    \end{tabular}%
    }
\end{table}
\begin{table}[t!]
    \centering
    \caption{Impacts on the system with load-altering attacks where PL = power loss}
    \label{tab:my-attack_power}
    \resizebox{\columnwidth}{!}{%
    \begin{tabular}{|c|c|c|}
    \hline
    \rowcolor{gray!30}
    \textbf{Load Increase} & \textbf{Real PL {[}kW{]}} & \textbf{Reactive PL {[}kVAr{]}} \\ \hline
    5\% & 209.45 & 139.80 \\ \hline
    10\% & 212.20 & 141.72 \\ \hline
    15\% & 214.97 & 143.66 \\ \hline
    \end{tabular}%
    }
\end{table}
Figure \ref{base_attack} illustrates the significant degradation in voltage profile under load-altering attacks, especially at 10\% and 15\% load increases, where the voltage falls below the acceptable range of 0.917 p.u. This highlights the vulnerability of the base case system to even modest cyberattacks targeting load distribution.

\subsection{Case Study 2: Impact of DG Integration on System Cybersecurity}

\begin{figure}[t!]
    \centering
   \includegraphics[width=0.5\textwidth]{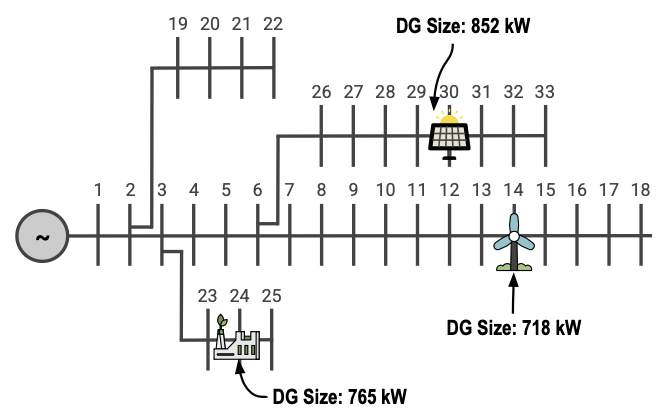} 
    \caption{IEEE 33-bus system layout with DGs} \label{fig:dg_ieee33} 
\end{figure}

\begin{figure}[t!]
    \centering
    \includegraphics[width=0.49\textwidth]{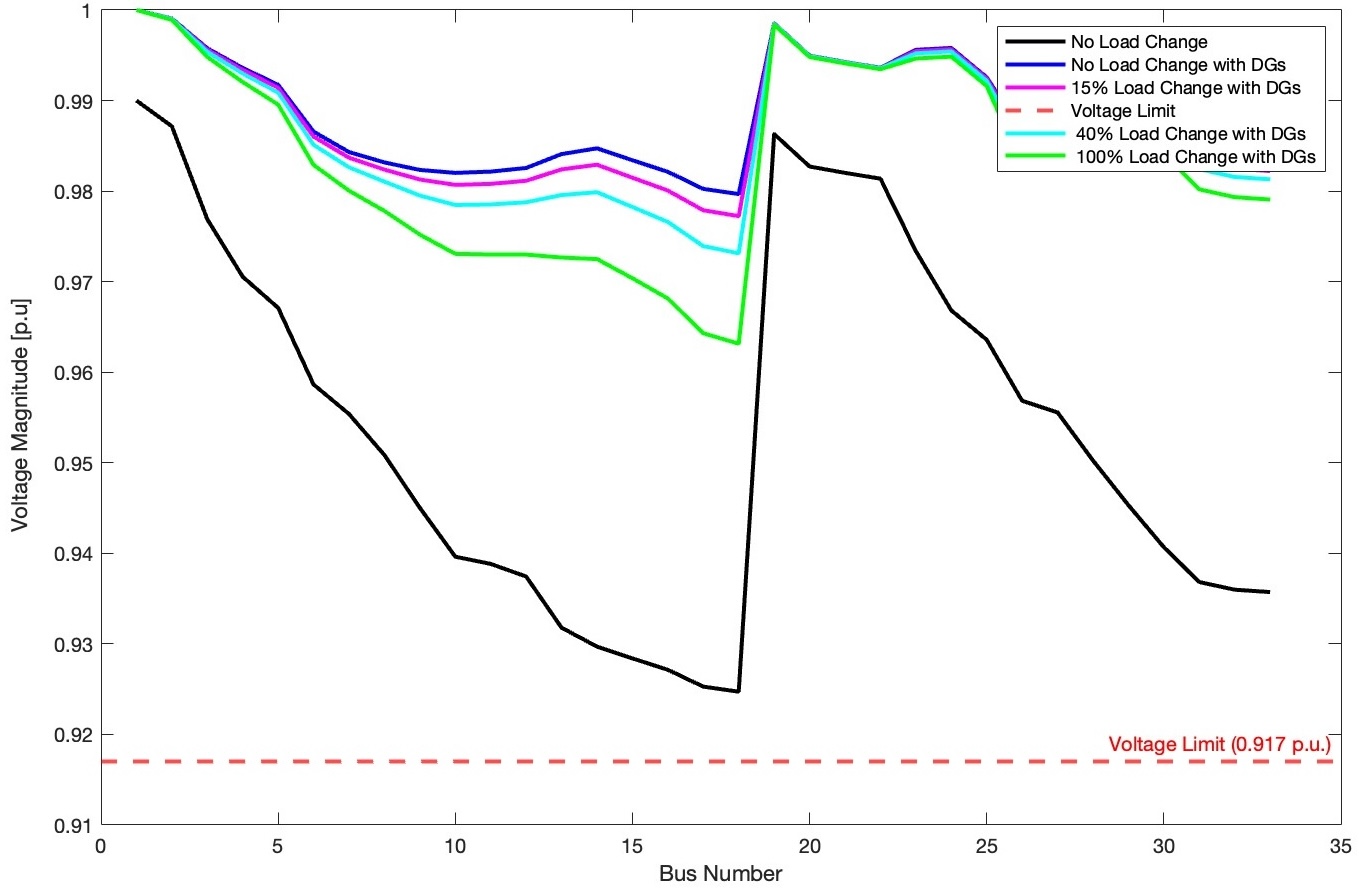}
    \caption{Voltage profile for DG-integrated system under step attacks}
    \label{fig:dg_vp}
\end{figure}

To enhance the resilience of the IEEE 33-bus system, the optimal placement and sizing of three DG units is determined using PSO. DGs are placed at Buses 30, 24, and 14 with capacities of 852 kW, 765 kW, and 718 kW, respectively, as shown in Figure \ref{fig:dg_ieee33}. These placements improve the voltage profiles from the 0.92 p.u. range to 0.98 p.u., as shown in Figure \ref{fig:dg_vp}, where the black line represents the base case and the blue line represents the system with DGs.
%
After DG placement, the critical buses are re-identified using VM, pinpointing Buses 10, 17, and 18 as the lowest voltage magnitudes. These buses are the new targets of the load-altering cyberattacks, with results summarized in Table \ref{tab:all_attacks} and Figure \ref{fig:dg_vp}. Under these conditions, the system remains stable under step load attacks, even with a 100\% load increase at five critical nodes, including Buses 11 and 16.
\begin{table*}[t!]
    \centering
    \caption{Impacts on the DG system with load-altering attacks}
    \label{tab:all_attacks}
    \resizebox{\textwidth}{!}{%
        {\normalsize
        \begin{tabular}{|c|c|c|c|c|c|}
        \hline
        \rowcolor{gray!30}
        \textbf{\begin{tabular}[c]{@{}c@{}}Attacked Nodes\end{tabular}} & \textbf{Number of DGs} & \textbf{\begin{tabular}[c]{@{}c@{}}Load Increase\end{tabular}} & \textbf{\begin{tabular}[c]{@{}c@{}}Real Power Loss\\ {[}kW{]}\end{tabular}} & \textbf{\begin{tabular}[c]{@{}c@{}}Reactive Power Loss\\ {[}kVAr{]}\end{tabular}} & \textbf{\begin{tabular}[c]{@{}c@{}}Lowest Voltage Magnitude \\ {[}p.u.{]}\end{tabular}} \\ \hline
        none      & 0 & 0\%   & 111.57 & 72.94 & 0.9247 \\ \hline
        none      & 3        & 0\%   & 29.66  & 21.41 & 0.9797 \\ \hline
        10, 17, 18  & 3     & 5\%   & 29.94  & 21.62 & 0.9789 \\ \hline
        10, 17, 18  & 3     & 10\%  & 30.23  & 21.85 & 0.9781 \\ \hline
        10, 17, 18  & 3     & 15\%  & 30.54  & 22.08 & 0.9772 \\ \hline
        10, 17, 18  & 3     & 40\%  & 32.29  & 23.42 & 0.9731 \\ \hline
        10, 17, 18   & 3    & 100\% & 37.96  & 27.80 & 0.9631 \\ \hline
        10, 16, 17, 18 & 3  & 50\%  & 34.15  & 24.79 & 0.9695 \\ \hline
        10, 16, 17, 18  & 3 & 100\% & 41.03  & 30.03 & 0.9587 \\ \hline
        10, 11, 16, 17, 18 & 3 & 50\%  & 35.28 & 25.60 & 0.9681 \\ \hline
        10, 11, 16, 17, 18 & 3 & 100\% & 43.95 & 32.12 & 0.9563 \\ \hline
        \end{tabular}%
        }
        }
\end{table*}
This result demonstrates that conventional DG placement strategies, initially designed to minimize losses and improve voltage profiles, can also enhance resilience to cyber-physical threats. DGs provide localized power support near load centers, reducing reliance on upstream sources and mitigating the impact of abrupt load changes. This decentralized structure acts as a natural ``shock absorber'' that mitigates step attack effects without the need for an explicit cybersecurity-aware design.

\subsection{Case Study 3: Impact of Dynamic Attacks}

\begin{figure}[t!]
    \centering
    \includegraphics[width=0.48\textwidth]{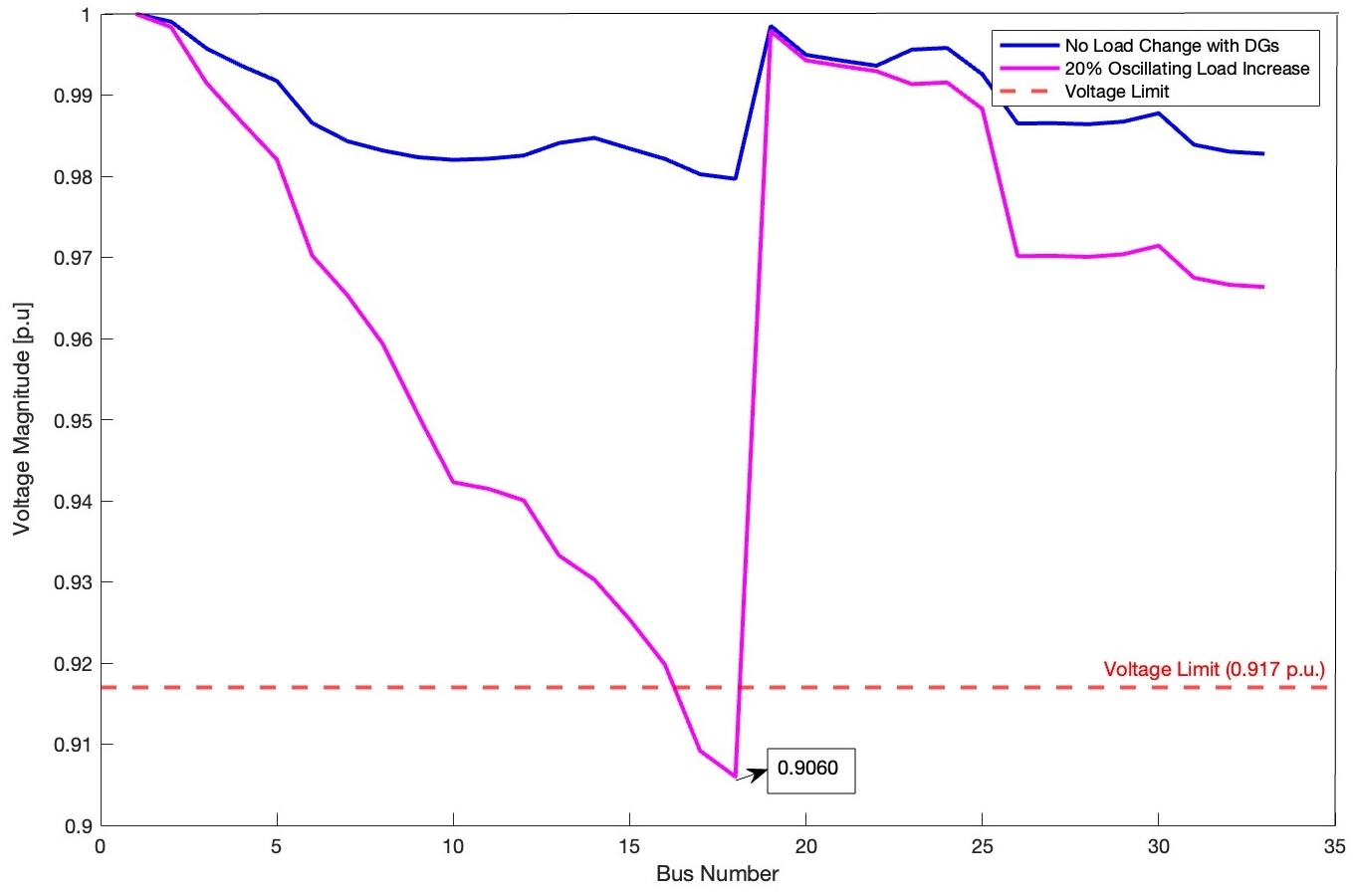}
    \caption{Voltage profile for DG-integrated system under 20\% oscillating attacks}
    \label{fig:osc_attack}
\end{figure}

\begin{figure}[t!]
    \centering
    \includegraphics[width=0.5\textwidth]{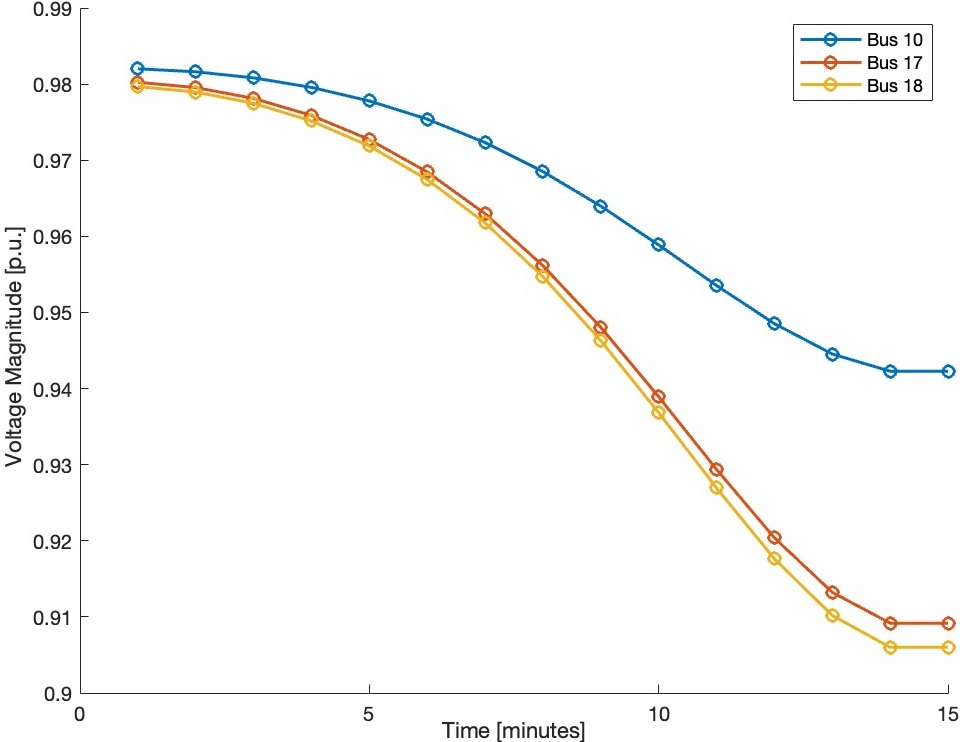}
    \caption{Impact of time-varying attack on critical node voltages}
    \label{fig:step_wise_attack}
\end{figure}

While step attacks offer a baseline for evaluating system resilience, real-world cyber threats are typically dynamic, gradually unfolding and varying over time. These attacks involve continuous manipulation of system parameters, such as loads or generation levels, and are harder to detect and mitigate. Unlike step attacks, which have immediate and observable effects, dynamic attacks are stealthier and more damaging due to their prolonged nature.
A 20\% load increase at critical nodes (Buses 10, 17, and 18), oscillating with a 50\% frequency component over 15 minutes, is modeled to simulate a dynamic attack. Though limited in scope, the Simulink simulation (Figure \ref{fig:osc_attack}) provides insights into the system's dynamic behavior under cyber-induced stress.
In the simulation, the attack involves ramped-up load injections at the identified critical nodes. The gradual load increases lead to voltage declines at Buses 10, 17, and 18 as shown in Figure \ref{fig:step_wise_attack}. Initially, DGs absorb the stress, maintaining voltage near nominal values. However, as the attack persisted, voltage stability begins to degrade, with the minimum voltage dropping to 0.906 p.u., below the 0.917 p.u. threshold used as a resilience benchmark.
%
This finding highlight the need to integrate cybersecurity-aware control strategies into grid design. Real-time anomaly detection, predictive modeling, and adaptive protection schemes should complement DG deployment to form a layered defense. Combining these tools with traditional optimization methods would enhance efficiency and protect against advanced persistent threats (APTs) exploiting temporal vulnerabilities.

\subsection{Discussion}
While DGs have demonstrated resilience against certain cyberattacks, this research represents just the starting point. Future studies should extend the attack replication methodology to larger, real-world distribution systems to better assess the scale and impact of cyber threats. Additionally, incorporating advanced stability indices such as VSI or FVSI will help evaluate the criticality of different nodes. Together, these will provide deeper insights into the severity of these threats on larger systems and determining if the impacts are more significant than those observed in this study.
However, it is clear that current DG allocation and sizing methods alone are no longer sufficient to address emerging cyber risks. This work lays the groundwork for developing more robust, cyber-aware grid designs that consider both operational efficiency and security. Multi-objective optimization models that integrate DG placement with real-time monitoring and adaptive grid management are essential to building more resilient systems. As smart technologies and renewable energy sources continue to integrate into the grid, ongoing research must continually reassess the role of DGs in mitigating evolving cyber threats.

\section{Conclusion}

This paper investigated the vulnerability of distribution systems to coordinated load-altering cyberattacks and proposed a resilience strategy based on optimal DG placement. Using the IEEE 33-bus distribution network as a testbed, we modeled MaDIoT-style attacks targeting voltage-sensitive nodes and evaluated their impact on voltage stability and system losses. Controlled load fluctuations of up to 15\% resulted in real and reactive power loss increases of 52\% and 51\%, respectively, at critical nodes in the baseline system. Following DG integration via PSO, voltage profiles were significantly improved. All buses maintained voltages above 0.97 p.u. even under extreme conditions such as a 100\% increase in load at critical nodes. These results demonstrate that DG optimization strategies, when applied with a resilience-aware mindset, can provide substantial 
resilience against cyber-physical threats.

\bibliographystyle{IEEEtran}
\bibliography{keylatex}

\end{document}